**To the Editor: comment on Eremets, M. I., Troyan, I. A., Nature Mater. 10, 927-931 (2011)**

Eremets and Troyan [1] claim that they produced the conducting liquid hydrogen state at 270 GPa and 295 K. Their evidence consists of disappearance of Raman signals, visual observations, and measurements of electrical conductivity in diamond anvil cells (DAC). However, there is no proof that the reported observations are due to transformations in hydrogen.

State-of-the-art observations of materials metallization under high pressures require measurements of the optical conductivity [2] and/or electrical resistivity [3]. The experiments of ref. 1 were performed using coated diamond anvils and the electrical leads were positioned at the opposite anvils. Optical properties of materials in the sample cavity performed though the coated diamonds are drastically affected by the presence of the interface layer even with the thickness of 10 nm (Fig. 1). The results of modeling of the optical properties show that in the experimental conditions of ref. 1, no conclusions can be made about the dc optical conductivity (connected to the electrical conductivity) based on observations of visible reflectance of the composite sample in the DAC cavity. The Raman data presented in ref. 1 report on disappearance of the signal of hydrogen above 260 GPa (cf. Ref. [4]). However, the presented data do not allow to judge whether the Raman peaks really disappear or become undetectable due to the increased background. Concerning the electrical conductivity measurements, we note that the electrical conductivity of hydrogen determined in ref. [1] is by at least by three orders of magnitude smaller than minimum value of metallic conductivity for hydrogen [5], and the contact resistance between the sample and the leads could not be singled out. Furthermore, the electrical contact geometry of ref. 1 is susceptible to uncontrollable shortening. We suggest that, as no data on the sample cavity thickness under pressure is presented in ref. 1, no experimental observations evidence is provided for the presence of hydrogen between the electrical leads in the sample cavity above 260 GPa.

Indeed, as hydrogen is squeezed out from the observation zone, changes in optical properties and electrical conductivity (shortening) would occur consistent with those reported in ref. 1. Moreover, thinning down of hydrogen sample would be consistent with the observations of disappearance of the Raman signal. The observed pressure hysteresis is consistent with this interpretation because diamond anvil cells are known to show pressure hysteresis effects due to plastic deformations of the gasket between the anvils. In conclusion, none of the data presented prove the existence of metallic liquid hydrogen under the conditions of claim.

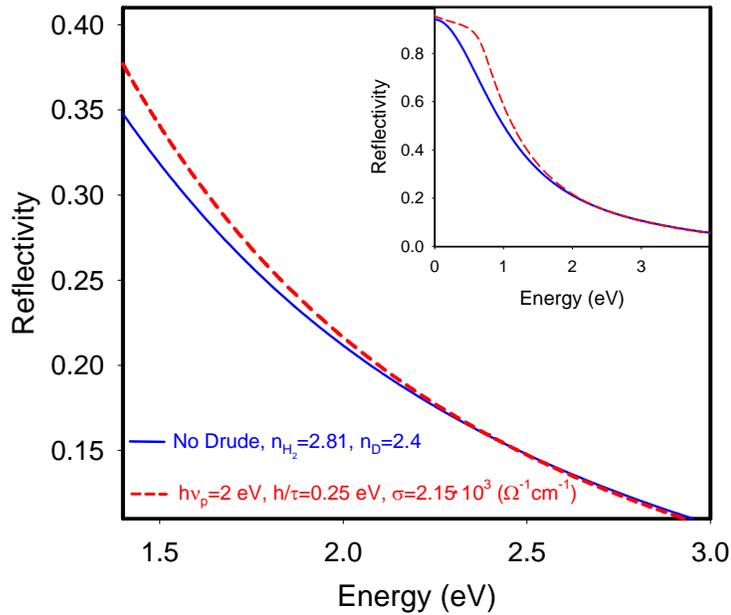

Fig. 1. Reflectivity of hydrogen sample in contact with a gold coated (10 nm thick) diamond anvil in the visible spectral range (inset: the same curves with IR and UV ranges included). Solid line -reflectivity of the insulating sample (no Drude contribution to conductivity); dashed line - reflectivity of the metallic sample, with the plasma frequency $\omega_p$=2 eV, the maximum value compatible with the optical absorption measurements of Ref. [4], the scattering rate $\gamma = h/\tau = 0.25$ eV is chosen to give the minimal value of metallic dc conductivity $\sigma$=2100 $\Omega^{-1}$cm$^{-1}$ determined in Ref. [5]. Refractive index of $H_2$, $n_{H2}$=2.81 is determined by extrapolating the results of measurements reported in Ref. [6] to 260 GPa; refractive index of diamond, $n_D$=2.4 (qualitatively results do not vary if other value of $n_D$ is used).

Alexander F. Goncharov, Viktor. V. Struzhkin
Geophysical Laboratory, Carnegie Institution of Washington